\UseRawInputEncoding 
\documentclass[pra,onecolumn,superscriptaddress]{revtex4}%
\usepackage{amssymb}
\usepackage{amsmath}
\usepackage{graphicx}
\usepackage{dcolumn}
\usepackage{xcolor}
\usepackage{braket}
\usepackage{bm}
\usepackage{subfigure}
\usepackage{amsfonts}
\usepackage{appendix}
\usepackage{tikz}
\usepackage[utf8]{inputenc}   
\usepackage[T1]{fontenc}      
\usepackage{lmodern}          
\usetikzlibrary{quantikz2,backgrounds,fit,decorations.pathreplacing}
\usepackage{xcolor}%
\usepackage{hyperref}
\def\path#1{#1}
\providecommand{\U}[1]{\protect\rule{.1in}{.1in}}

\renewcommand{\Im}{{\rm Im}}

\begin{document}
\preprint{APS/123-QED}

\title{Let the Qudit Do the Jacobi:
A Structured Quantum Algorithm for Spectral Decomposition}

\author{ Aikaterini Mandilara}

\affiliation{Department of Informatics and Telecommunications, National and Kapodistrian
University of Athens, Panepistimiopolis, Ilisia, 15784, Greece}
\affiliation{Eulambia Advanced Technologies Ltd, Agiou Ioannou 24, Building Complex C, Ag. Paraskevi, 15342, Greece}

\begin{abstract}
Jacobi diagonalization is a long-established numerical algorithm for the spectral decomposition of Hermitian and, more generally, normal matrices. In this work, we develop a qudit-native quantum realization of the Jacobi diagonalization algorithm for unknown unitary operators. The proposed framework avoids explicit reconstruction of the operator and controlled-unitary operations. Each elementary two-parameter Givens rotation is implemented through two sequential single-parameter quantum variational optimizations that are directly accessible experimentally. An interferometric protocol is further introduced for extracting the eigenvalues of the diagonalized unitary, up to an overall global phase. Numerical simulations on ensembles of Haar-random unitary matrices demonstrate that the proposed algorithm preserves the characteristic convergence behavior of the classical  Jacobi method and exhibits the expected quadratic scaling in the number of elementary operations with the dimension $d$, comparable to its classical counterpart. The results establish the proposed algorithm as a structured quantum numerical linear algebra algorithm naturally suited to qudit architectures and provide a bridge between classical iterative matrix algorithms and their quantum realizations.
\end{abstract}

\maketitle

\section{Introduction \label{intro}}


Spectral decomposition is a fundamental concept in mathematics and physics, while at the same time playing a central role in numerous applications in engineering and computer science. Quantum physics is no exception, as many theoretical and algorithmic problems ultimately require the determination of the spectrum and eigenvectors of either a Hamiltonian or a unitary operator. Such tasks arise both in the study of quantum systems and in the development of quantum algorithms.

Broadly speaking, two different approaches can be followed for the diagonalization of an unknown quantum operator. In the first, the quantum operator is converted into classical information through quantum process tomography \cite{Chuang1997, NielsenChuang, Nielsen2021gatesettomography}, after which conventional numerical linear algebra techniques are employed to obtain its spectral decomposition. In the second approach, one retains the quantum nature of the operator throughout the computation and attempts to drive it coherently towards a diagonal form, extracting the eigenvalues only at the final stage. The latter philosophy treats the quantum operator as a native quantum object rather than first converting it into a classical matrix representation. This constitutes the focus of the present work.


Several quantum algorithms have been proposed for coherent eigensystem extraction. Among the earliest and most influential is the Quantum Phase Estimation (QPE) algorithm,  introduced in \cite{Kitaev1995} and used as a central component of Shor's factoring algorithm \cite{Shor1994}. Assuming that an eigenvector of the unitary operator is available, QPE estimates the corresponding eigenphase to arbitrary  precision. Despite its remarkable asymptotic performance, practical implementations remain demanding, as they require controlled-unitary operations together with deep coherent circuits. While several variants of QPE have subsequently been proposed, including iterative \cite{Dobsicek2007}
and Bayesian implementations \cite{Wiebe2016},
their experimental realization remains demanding.

More recently, variational and hybrid quantum-classical algorithms have emerged as an attractive approach for noisy intermediate-scale quantum devices. Beyond Variational Quantum Eigensolvers \cite{Peruzzo2014} for estimating low-energy eigenstates, several variational frameworks have been proposed for the diagonalization and decomposition of quantum operators, including Variational Quantum State Diagonalization \cite{Nakanishi2019VQSD} and Variational Quantum Singular Value Decomposition \cite{Wang2021VQSVD}. These methods formulate spectral decomposition as a global variational optimization problem over parameterized quantum circuits. In contrast, the present work follows a structured numerical linear algebra approach, realizing the classical generic Jacobi diagonalization procedure through a sequence of local quantum operations.

A rather different philosophy originates from classical numerical linear algebra. Instead of treating eigensystem extraction either through coherent phase estimation or through global variational optimization, structured matrix algorithms progressively eliminate the off-diagonal elements by a sequence of local similarity transformations. Among these methods, Jacobi diagonalization \cite{Jacobi1846} occupies a distinguished position owing to its conceptual simplicity, robustness, and well-understood convergence properties. Since its original introduction,  the method \cite{GolubVanLoan2013} has been generalized from real symmetric matrices to Hermitian, unitary and, more generally, normal matrices \cite{Goldstine1959}, while preserving many of its favorable convergence characteristics \cite{Ruhe1967,Loizou1971}. Surprisingly, despite this extensive classical development, comparatively little attention has been devoted to realizing the Jacobi diagonalization procedure itself as a quantum algorithm acting directly on unknown quantum operators.


In the present work, we develop a qudit-native realization of the Jacobi diagonalization procedure, which we henceforth refer to as the Jacobi Qudit Algorithm (JQA). The key observation is that the elementary two-level Givens rotations underlying the classical Jacobi algorithm naturally correspond to $SU(2)$ subgroups of $SU(d)$ and can therefore be  implemented within a single qudit. As discussed later in Sec.~\ref{discussion}, this correspondence becomes considerably less direct in qubit architectures, where the same transformations generally require compilation into sequences of elementary one- and two-qubit gates.

Beyond their natural suitability for JQA, qudits have recently attracted considerable attention as a computational platform \cite{Wang,Deller,Weggemans}. Their larger Hilbert space has been exploited in applications ranging from high-dimensional quantum communication \cite{Cozzolino,Luo} and quantum error correction \cite{Byrd2005,Brock2025} to quantum machine learning \cite{Wach,Qudit,Maragkopoulos2025}, while numerous experimental platforms—including photonic, trapped-ion, neutral-atom, and superconducting architectures—have demonstrated increasingly mature control of higher-dimensional quantum systems (see, for instance, \cite{Erhard,Kues,Ringbauer,Blok}). These developments make qudits an appealing platform for implementing structured quantum numerical algorithms such as the one proposed here.


Based on these observations, the main contributions of the present work are summarized as follows:
\begin{enumerate}
    \item We propose a structured quantum algorithm for the diagonalization of an unknown unitary operator acting on a single qudit. The algorithm follows the philosophy of the classical Jacobi method while avoiding explicit reconstruction of the matrix.

    \item We demonstrate that each elementary two-parameter Jacobi rotation can be replaced by two sequential single-parameter variational optimizations that naturally map onto experimentally accessible qudit operations, while preserving the convergence behavior of the underlying Jacobi procedure.

    \item We introduce an interferometric protocol for extracting the eigenvalues of the diagonalized unitary operator, up to an overall global phase, thereby completing the diagonalization procedure entirely within the quantum domain.

    \item Through numerical simulations on Haar-random unitary matrices, we verify convergence under cyclic pivoting, demonstrate an approximately constant number of sweeps with increasing dimension, and show an empirical $\mathcal{O}(d^2)$ scaling in the total number of pivot operations required to reach a prescribed accuracy.

    \item Finally, we discuss several practical aspects of the proposed framework, including restricted connectivity, acceleration strategies, the relation between qudit and qubit implementations, and its potential relevance to early fault-tolerant quantum computing.
\end{enumerate}


The remainder of the manuscript is organized as follows. In Sec.~\ref{Background}, we review the classical Jacobi diagonalization algorithm together with the main theoretical ingredients required for the developments presented in this work. The same section also introduces the algebraic framework of qudits and the relevant properties of the $\mathfrak{su}(d)$ generators employed throughout the manuscript.
In Sec.~\ref{Jacobi}, we present the proposed  algorithm --JQA, which consists of three main components: \textit{(a)} the iterative Jacobi diagonalization scheme, \textit{(b)} the local single-parameter variational optimization used to determine the elementary rotations, and \textit{(c)} the interferometric extraction of the eigenvalues upon completion of the diagonalization procedure. We also discuss the expected convergence properties and the computational complexity of JQA.
Section~\ref{numerics} presents numerical experiments on ensembles of Haar-random unitary matrices, validating the convergence behavior and empirical scaling of the algorithm. Finally, Sec.~\ref{discussion} discusses practical implementation aspects, including acceleration strategies, restricted connectivity, qudit-to-qubit mappings and relevance to early fault-tolerant architectures.

\section{Background \label{Background}}

In this section, we review the background material required for the development of the JQA. We first summarize the classical Jacobi diagonalization procedure for normal matrices and subsequently introduce the elements of qudit quantum mechanics that are used throughout the remainder of the manuscript.

\subsection{Classical generic Jacobi diagonalization of normal matrices}

The Jacobi algorithm is a classical iterative method for the diagonalization of real symmetric matrices through successive applications of Givens rotations, originally introduced by Jacobi in 1846 \cite{Jacobi1846}. Owing to its conceptual simplicity, numerical robustness, and favorable convergence properties, the method has subsequently been extended to broader classes of matrices. In this work, we employ the extension to normal matrices, i.e. matrices satisfying $M^{\dagger}M = MM^{\dagger}$, commonly referred to as the Jacobi algorithm for normal matrices \cite{Ruhe1967}. Since this algorithm constitutes the basis of the proposed quantum framework, we briefly summarize its main steps below.

Let $M \in \mathbb{C}^{d \times d}$ be a normal matrix. The algorithm relies on unitary transformations acting nontrivially on two-dimensional subspaces. For each  pair of indices $(p,q)$ satisfying $1 \leq q < p \leq d$, we define a family of unitary Givens rotations $U^G_{(p,q)}(\phi,\alpha)$ acting on the $(p,q)$-plane. The matrix elements of $U^G_{(p,q)}$ are given by
\begin{align}
    [U^G_{(p,q)}]_{i,j} &= \delta_{i,j}, \qquad (i,j) \neq \{(p,p),(q,q),(p,q),(q,p)\}, \nonumber\\
    [U^G_{(p,q)}]_{p,p} &= [U^G_{(p,q)}]_{q,q} = \cos\phi, \nonumber\\
    [U^G_{(p,q)}]_{q,p} &= e^{i\alpha} \sin\phi, \nonumber\\
    [U^G_{(p,q)}]_{p,q} &= -e^{-i\alpha} \sin\phi .\label{Upq}
\end{align}
These matrices are unitary for all real parameters $\phi$ and $\alpha$.

The infinitesimal generators associated with unitary rotations in the $(p,q)$ subspace are given by the off-diagonal generators of the Lie algebra $\mathfrak{su}(d)$,
\begin{align}
    \hat{g}^{(R)}_{(p,q)} &= \ket{p}\bra{q} + \ket{q}\bra{p}, \nonumber\\
    \hat{g}^{(I)}_{(p,q)} &= -i\ket{p}\bra{q} + i\ket{q}\bra{p}, \label{gener}
\end{align}
with  $1 \leq q < p \leq d$ and where $\ket{p}$ denotes the $p$-th canonical basis vector in $\mathbb{C}^d$ and $\bra{p}$ its Hermitian adjoint. The unitary Givens rotations act nontrivially only on the two-dimensional subspace spanned by $\{\ket{p},\ket{q}\}$ and one can express the matrix described in Eq.(\ref{Upq}) as 
\begin{equation}
    U^G_{(p,q)}=\exp\left[ i \phi(\cos\alpha ~\hat{g}^{(R)}_{(p,q)}+\sin\alpha ~\hat{g}^{(I)}_{(p,q)}) \right].\label{UGpq}
    \end{equation}

A unitary similarity transformation
\begin{equation}
    \tilde{M} = U M U^{\dagger}
\end{equation}
preserves the spectrum of $M$.  Since the transformation is unitary, it also preserves the Frobenius norm
\begin{equation}
    \|M\|_F^2 = \sum_{i,j} |M_{i,j}|^2 .
\end{equation}
Applying Givens rotation $U^G_{(p,q)}$ yields
\begin{equation}
    \tilde{M} = U^G_{(p,q)} M U^{G\dagger}_{(p,q)} ,
\end{equation}
which leaves invariant both the Frobenius norm of the full matrix and that of the corresponding $(p,q)$ sub-block that we denote as,  $\|M_{(p,q)}\|_F^2$. 

For normal matrices, it can be shown that suitable choices of the parameters $(\phi,\alpha)$ allow one to decrease the quantity
\begin{equation}
    \tau_{(p,q)}(M) = |M_{p,q}|^2 + |M_{q,p}|^2 , \label{locost}
\end{equation}
while  $\|M_{(p,q)}\|_F^2$ is preserved. Defining the global off-diagonal cost function
\begin{equation}
    \tau(M) = \sum_{i \neq j} |M_{i,j}|^2 = \|M\|_F^2 - \sum_j |M_{j,j}|^2 , \label{glocost}
\end{equation}
one can construct an iterative procedure based on successive Givens rotations that monotonically decreases $\tau(M)$. It is shown in \cite{Ruhe1967,Loizou1971} that $\tau(M)$ converges to zero with a number of Givens rotations scaling as $\mathcal{O}(d^2)$. Since unitary similarity transformations preserve normality, the vanishing of the off-diagonal cost function implies convergence to a diagonal matrix.

The algorithm can be described by considering a generic iteration $k$, starting from  $M^{(0)} = M$: 
\begin{enumerate}
    \item Select a pair of indices $(p,q)$ corresponding to a $2\times2$ sub-block of $M^{(k-1)}$, according to a prescribed pivoting strategy (e.g. maximal-element or cyclic pivoting).
    \item Apply a unitary Givens rotation
    \begin{equation}
    M^{(k)} = U^G_{(p,q)} M^{(k-1)} U^{G\dagger}_{(p,q)} .
        \end{equation}
    \item Choose the parameters $(\phi,\alpha)$ so as to minimize $\tau_{(p,q)}(M^{(k)})$.
\end{enumerate}

 In contrast to the Jacobi algorithm for real symmetric matrices, where the rotation angle can be  determined analytically  to exactly annihilate the off-diagonal entry, the minimization for normal matrices involves coupled nonlinear relations between the two rotation parameters and admits a non-unique family of minimizing solutions. In addition, $\tau_{(p,q)}$ cannot, in general, be reduced exactly to zero.

\subsection{Qudit systems}

A qudit is the natural generalization of a qubit from a two-dimensional to a $d$-dimensional quantum system.
More specifically, a qudit state resides in a $d$-dimensional Hilbert space equipped with a chosen orthonormal basis. We denote this basis as $\{\ket{j}\}_{j=0}^{d-1}$ and use it as the computational basis of the qudit. A general pure qudit state can then be expressed as
\begin{equation}
\ket{\psi} = \sum_{j=0}^{d-1} c_j \ket{j},
\end{equation}
where the complex coefficients $c_j$ satisfy the normalization condition $\sum_{j=0}^{d-1} |c_j|^2 = 1$.

In this work, we consider the full $\mathfrak{su}(d)$ Lie algebra associated with the qudit system, which is spanned by $d^2 - 1$ traceless Hermitian generators. This set includes the $d(d-1)$ off-diagonal generators, introduced in the previous subsection (see Eqs.~(\ref{gener})), as well as diagonal generators. The latter may be chosen, for each $l$ with $0 \leq l \leq d - 2$, as
\begin{equation} 
\hat{g}^{(D)}_l = \sqrt{\frac{2}{(l+2)(l+1)}} \left( \sum_{j=0}^{l} \ket{j}\bra{j} - (l+1) \ket{l+1}\bra{l+1} \right).
\end{equation}

In the following we assume the existence of a Hermitian observable $B$, diagonal in the computational basis and possessing a non-degenerate spectrum. Such an observable can always be constructed as an appropriate linear combination of the diagonal generators $\hat{g}^{(D)}_l$. Its spectral decomposition is
\begin{equation}
    B = \sum_{j=0}^{d-1} \beta_j \ket{j}\bra{j}  \label{B}
\end{equation}
where the eigenvalues $\{\beta_j\}$ are real and pairwise distinct.

\section{The Jacobi Qudit Algorithm}\label{Jacobi}

The aim of this section is to develop a quantum algorithm for the diagonalization of an unknown unitary operator (or, equivalently, the unitary evolution generated by a Hamiltonian) acting on a single qudit, including the extraction of its eigenvalues and the construction of its eigenvectors.  The proposed iterative procedure closely follows the structure of the generic  Jacobi algorithm. However, we adopt an alternative parametrization of unitary Givens rotations, which leads to a different and more practical strategy for identifying the optimal rotations on quantum hardware.

In particular, the required Jacobi rotation parameters are determined experimentally through successive one-dimensional variational searches. The resulting circuit implementation requires only repeated measurements of the observable B, defined in Eq.~(\ref{B}), performed at the end of the circuit.

We begin by describing the iterative update rule that forms the core of the algorithm. The subsequent subsections present the variational determination of the optimal Givens rotations, the extraction of eigenvalues and eigenvectors, and finally the convergence properties and computational complexity of JQA.

\subsection{The iteration scheme}

We begin by introducing two one-parameter families of unitary operations acting non-
trivially on the two-dimensional subspace spanned by $\{\ket{p}, \ket{q}\}$:
\begin{align}
     U^R_{(p,q)}(\phi_1) &= \exp\!\left[ i \phi_1 \, \hat{g}^{(R)}_{(p,q)} \right], \nonumber\\
     U^I_{(p,q)}(\phi_2) &= \exp\!\left[ i \phi_2 \, \hat{g}^{(I)}_{(p,q)} \right], \label{UGRI}
\end{align}
which we refer to as the \emph{real} and \emph{imaginary} unitary Givens rotations, respectively. Each of these transformations acts as the identity on all basis states except $\ket{p}$ and $\ket{q}$.

We now describe the iterative procedure for diagonalizing an unknown unitary operator $M$ acting on the qudit. For concreteness, we consider a cyclic pivoting strategy, in which the index pairs $(p,q)$ are chosen sequentially in a predetermined order. The procedure is schematically depicted in Fig.~(\ref{fig1})(a).

Let $M^{(0)} = M$. At the $k$-th iteration, corresponding to a fixed pair $(p,q)$, the following steps are performed:
\begin{enumerate}
    \item Apply a real unitary Givens rotation via conjugation,
    \begin{equation}
         R(\phi_1)=U^R_{(p,q)}(\phi_1) \, M^{(k-1)} \, U^{R\dagger}_{(p,q)}(\phi_1) ,
    \end{equation}
    and identify, through a quantum variational search (see Sec.~\ref{QVC}), the angle $\tilde{\phi}_1$ that minimizes the local cost function $\tau_{(p,q)}(R(\phi_1))$. This step requires repeated measurements of the observable $B$.

    \item Update, $U^R_{(k)}=U^R_{(p,q)}(\tilde{\phi}_1)$ and $M^{(k)}_R=U^R_{(k)} \, M^{(k-1)} \, U^{R\dagger}_{(k)}$.

    \item Apply an imaginary unitary Givens rotation,
    \begin{equation}
      I(\phi_2)= U^I_{(p,q)}(\phi_2) \, M^{(k)}_R \, U^{I\dagger}_{(p,q)}(\phi_2) ,
    \end{equation}
    and, again via quantum variational search, determine an angle $\tilde{\phi}_2$ that minimizes $\tau_{(p,q)}( I(\phi_2))$.

    \item Update, $U^I_{(k)}=U^I_{(p,q)}(\tilde{\phi}_2)$ and $M^{(k)}=U^I_{(k)} \, M^{(k)}_R \, U^{I\dagger}_{(k)}$.
\end{enumerate}

At this point it is instructive to compare the present construction with the  generic Jacobi algorithm. In the classical formulation, one could combine the above steps into a single optimization over two parameters by applying the unitary Givens rotation Eq.(\ref{UGpq}). Here, however, we exploit the fact that the minimization problem admits a family of solutions and decompose it into two sequential single-parameter searches. As demonstrated by the numerical results, this decomposition is not only more convenient for experimental implementation but also exhibits improved convergence compared to the generic Jacobi algorithm.

\subsection{One-dimensional quantum variational search\label{QVC}}

At each step of the algorithm, two independent one-dimensional variational optimizations are performed in order to identify the parameters $\phi_1$ and $\phi_2$ that minimize the local cost functions $\tau_{(p,q)}(R(\phi_1))$ and $\tau_{(p,q)}(I(\phi_2))$, respectively. Since the two optimization procedures are technically identical, we present them in a unified manner.

Let $A$ denote a unitary operator and define the one-parameter family
\begin{equation}
    A(\phi) = e^{i\phi \hat{g}_{(q,p)}} \, A \, e^{-i\phi \hat{g}_{(q,p)}},
\end{equation}
where $\hat{g}_{(q,p)}$ is one of the generators introduced in Eq.~(\ref{gener}). The variational task consists of minimizing the local cost function $\tau_{(p,q)}(A(\phi))$ over the bounded domain $-\pi \le \phi < \pi$. The resulting optimization landscape is smooth, and our numerical results indicate the absence of pathological features for dimensions up to $d \le 30$.

If one attempts to employ a gradient-based optimization method, the derivative of the cost function with respect to $\phi$ can be expressed as
\begin{equation}
\frac{d \tau_{(p,q)}(A(\phi))}{d\phi}
= 2 \, \Im \!\left[
\big( [A(\phi)]_{p,p} - [A(\phi)]_{q,q} \big)
\big( [A(\phi)]^{*}_{p,q} - [A(\phi)]^{*}_{q,p} \big)
\right],
\label{grad}
\end{equation}
which requires knowledge of all four complex elements of the $(p,q)$ sub-block. As a consequence, direct gradient estimation is experimentally demanding.  Furthermore, simple parameter-shift rules do not appear to apply directly to this cost function.

In contrast, the cost function itself can be efficiently estimated using only measurements of absolute squares of matrix elements. Suppose the qudit is initialized in the computational basis state $\ket{p}$ and define
\begin{equation}
    \ket{\psi} = A(\phi) \ket{p},
\end{equation}
where the parameter $\phi$ is fixed throughout the measurement procedure. Repeated measurements of the observable $B$ yield estimates of the probabilities
\begin{equation}
    |\braket{q | \psi}|^2
    = |\braket{q | A(\phi) | p}|^2
    = |[A(\phi)]_{q,p}|^2,
\end{equation}
for $q = 0, \ldots, d-1$, providing access to the absolute squares of the elements of the $p$-th column of $A(\phi)$. Repeating the procedure with the initial state $\ket{q}$ allows estimation of the corresponding elements of the $q$-th column. In this way, all quantities required to evaluate the local cost function $\tau_{(p,q)}(A(\phi))$ are obtained.

Motivated by these observations, we adopt a derivative-free one-dimensional variational search, illustrated in
Fig.~(\ref{fig1})(b), based solely on repeated evaluations of the local cost function. This strategy avoids explicit reconstruction of the operator while remaining naturally compatible with experimental implementations.

\subsection{Extracting the eigenvalues and eigenvectors}

Assuming that the JQA is implemented using a cyclic pivoting strategy, then a single Jacobi sweep consists of $d(d-1)/2$ steps, each involving two applications of a single-parameter quantum variational procedure. At the end of each sweep, the measurement procedure previously used to estimate the local cost function $\tau_{(p,q)}$ can be extended to estimate the global cost function $\tau(M)$ defined in Eq.~(\ref{glocost}).

Specifically, the qudit is initialized in each computational basis state $\ket{j}$, $j=0,\ldots,d-1$, and the observable $B$ is measured repeatedly. The resulting statistics provide estimates of the absolute squares of all matrix elements of the transformed operator $M^{(k)}$, allowing one to evaluate the global cost function. This provides a natural stopping criterion based on a prescribed threshold for the global cost function.

Let $K$ denote the final iteration index. The cumulative effect of the algorithm can be written as
\begin{align}
M^{(K)} &= U^I_{(K)} U^R_{(K)} \cdots U^I_{(1)} U^R_{(1)} \,
M \,
U^{R\dagger}_{(1)} U^{I\dagger}_{(1)} \cdots U^{R\dagger}_{(K)} U^{I\dagger}_{(K)} \nonumber \\
&= U_K \, M \, U_K^\dagger ,
\end{align}
where $U_K$ denotes the known unitary circuit generated by the algorithm. Upon convergence, $M^{(K)}$ is approximately diagonal, with diagonal elements corresponding to the eigenvalues of $M$, and the states $U_K \ket{j}$ providing the associated eigenvectors.

 We now address the problem of extracting the corresponding eigenvalues, assuming that
\begin{equation}
M^{(K)} \approx \mathrm{diag}(e^{i\phi_0}, \ldots, e^{i\phi_{d-1}}),
\end{equation}
where $\phi_j$ are the eigenphases of the unitary $M$. To extract the eigenphases, we employ interference between pairs of computational basis states. For this purpose we introduce the Hadamard-like unitary acting on the two-dimensional subspace spanned by $\{\ket{i},\ket{j}\}$ and as identity elsewhere:
\begin{equation}
H_{(i,j)} = \frac{1}{\sqrt{2}}
\left(
\ket{i}\bra{i} + \ket{i}\bra{j} + \ket{j}\bra{i} - \ket{j}\bra{j}
\right)+\sum_{k\ne i, j}\ket{k}\bra{k}.
\end{equation}

Preparing the qudit in the state $\ket{i}$, applying $H_{(i,j)}$, followed by $M^{(K)}$ and $H_{(i,j)}$ again, yields a final state whose measurement in the computational basis gives outcome $\ket{i}$ with probability
\[
P_i = \frac{1}{2}\big(1 + \cos(\phi_i - \phi_j)\big),
\]
and outcome $\ket{j}$ with probability $\frac{1}{2}(1 - \cos(\phi_i - \phi_j))$.

\begin{figure}[htbp]
\centering\includegraphics[width=15
cm]{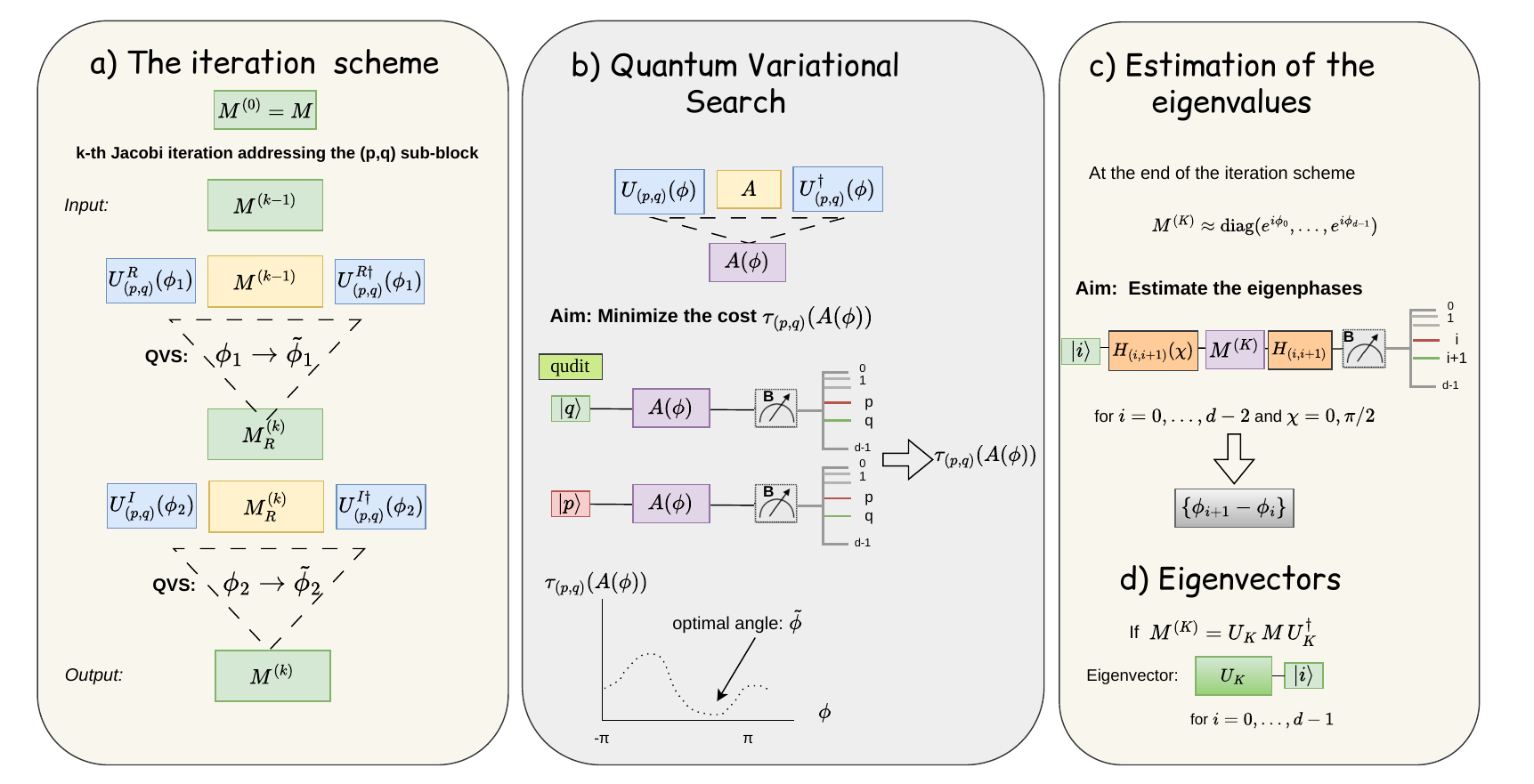}
\caption{Overview of the proposed Jacobi Qudit Algorithm (JQA). \textit{(a)} Iteration scheme for a single Jacobi step acting on the $(p,q)$ subspace. Each iteration consists of two successive one-dimensional quantum variational searches (QVS), determining the optimal parameters of the real and imaginary Givens rotations, followed by the corresponding updates of the transformed operator. \textit{(b)} Quantum variational search used to determine each rotation parameter. The local cost function is evaluated through repeated measurements of the observable $B$ on the computational basis states $\ket{p}$ and $\ket{q}$, and the optimal rotation angle $\tilde{\phi}$ is obtained by minimizing the resulting one-dimensional cost landscape. \textit{(c)} Eigenphase estimation after convergence of the Jacobi iteration. Hadamard-like interference between neighboring computational basis states allows the determination of successive phase differences, from which the complete spectrum is reconstructed up to an overall global phase. \textit{(d)} Construction of the eigenvectors. Once the algorithm has converged, the accumulated unitary circuit $U_K$
 maps each computational basis state onto the corresponding eigenvector of the original unitary operator.
 \label{fig1}}
\end{figure}

Repeating this procedure for adjacent index pairs $(i,i+1)$ allows one to estimate the phase differences $\phi_i - \phi_{i+1}$ up to a sign ambiguity. This ambiguity can be resolved by replacing the first Hadamard-like operation with a phase-shifted version,
\begin{equation}
H_{(i,j)}(\chi) = \frac{1}{\sqrt{2}}
\left(
\ket{i}\bra{i} + e^{i\chi}\ket{i}\bra{j}
+ \ket{j}\bra{i} - e^{i\chi}\ket{j}\bra{j}
\right)+\sum_{k\ne i, j}\ket{k}\bra{k}, \label{hachi}
\end{equation}
which modifies the interference term to $\cos(\phi_i - \phi_j + \chi)$. Choosing, for example, $\chi = 0$ and $\chi = \pi/2$  allows the sign of the phase difference to be determined. The procedure is summarized in Fig.~(\ref{fig1})~(c).

Fixing a reference phase (e.g., $\phi_0 = 0$), the full spectrum of eigenphases can thus be reconstructed. If the unitary $M$ is generated as $M = e^{-iHt}$ for a Hamiltonian $H$, this procedure provides access to the spectrum of $H$ up to an overall energy shift.

\subsection{Convergence and scaling \label{convergence}}

For the generic Jacobi algorithm reviewed in Section~\ref{Background}, it is known that the method converges globally with a linear rate in the number of applied unitary Givens rotations and exhibits quadratic convergence in the vicinity of the diagonal fixed point. Both numerical and analytical studies indicate that the number of Jacobi sweeps required to reach a prescribed accuracy remains approximately constant as the matrix dimension $d$ increases. Since each sweep consists of $d(d-1)/2=\mathcal{O}(d^2)$ Givens rotations and each rotation requires $\mathcal{O}(d)$ arithmetic operations, the overall computational complexity of the classical algorithm scales as $\mathcal{O}(d^3)$.

The proposed JQA differs from the classical generic Jacobi algorithm in only two respects:
\begin{itemize}
    \item each two-parameter Givens rotation, Eq.~(\ref{UGpq}), is replaced by two successive single-parameter rotations, Eq.~(\ref{UGRI});
    \item the corresponding rotation parameters are determined sequentially through two independent one-dimensional quantum variational searches.
\end{itemize}

Although a formal convergence proof for this modified procedure is presently unavailable, its close structural similarity to the classical algorithm suggests that the essential convergence properties are preserved. In particular, the two sequential optimizations explore the same two-dimensional rotation manifold associated with each pivot. The numerical results presented in Section~\ref{numerics} support this expectation, indicating convergence behavior consistent with that of the classical generic Jacobi method. Since the number of sweeps remains approximately constant, the total number of Jacobi iterations in JQA also scales empirically as $\mathcal{O}(d^2)$.

The above discussion concerns only the algorithmic structure of JQA. It does not include implementation-specific quantum resources, such as measurement-shot complexity or the cost of the variational searches, which are discussed in the following subsection.

\subsection{Quantum resource requirements}

The convergence analysis presented above characterizes the algorithmic scaling of the proposed JQA. Here we summarize the quantum resources required for its implementation, including state preparation, elementary operations, circuit depth, measurements, and the eigenvalue extraction procedure.

The algorithm requires the ability to initialize the qudit in arbitrary computational basis states and to implement the two families of parametrized two-level unitary operations introduced in Eqs.~(\ref{UGRI}). In addition, the eigenvalue extraction procedure requires the phase-shifted Hadamard-like operations of Eq.~(\ref{hachi}). Throughout the algorithm only a single non-degenerate observable $B$, Eq.~(\ref{B}), is measured, while neither ancilla qudits nor controlled-unitary operations are required.

Each Jacobi sweep consists of $\mathcal{O}(d^2)$ successive two-level qudit rotations. Since the number of sweeps required for convergence is observed numerically (see Section~\ref{numerics}) to remain approximately constant, the \textit{depth} of the accumulated circuit implementing the final unitary transformation $U_K$ scales empirically as $\mathcal{O}(d^2)$.

The dominant experimental cost originates from the repeated evaluation of the local cost function during the variational searches. At each Jacobi step, however, only the two columns (or equivalently the two rows) associated with the current pivot need to be accessed through measurements of the observable $B$. The overall measurement effort therefore depends on the desired statistical precision and on the optimization strategy employed, while remaining polynomial in the system dimension.

\section{Numerical Experiments\label{numerics}}

In this section, we investigate the convergence properties and empirical scaling of JQA through numerical simulations. The algorithm is applied to ensembles of Haar-random unitary matrices (see \cite{github}) using cyclic pivoting throughout. Convergence is monitored using two metrics. The first is the global off-diagonal cost function $\tau(M)$ , Eq.~(\ref{glocost}), while the second is the average eigenvalue error,
\begin{equation}
\epsilon=\frac{1}{d}\sum_{i=1}^{d}\left|\lambda_i-M_{i,i}\right|,\label{epsilon}
\end{equation}
where $\{\lambda_i\}$ denote the exact unimodular eigenvalues of the target unitary matrix and $M_{i,i}$ are the diagonal entries produced by the algorithm.

The numerical experiments address two questions. First, they examine the convergence behavior of the algorithm and the scaling of the number of Jacobi sweeps with the dimension $d$. Second, they compare the proposed sequential single-parameter implementation with the standard unitary Jacobi algorithm.

The simulations show that the sequential single-parameter implementation preserves the convergence behavior of the standard unitary Jacobi algorithm while often requiring slightly fewer sweeps to reach a prescribed accuracy.


\subsection{Convergence under cyclic pivoting}

\begin{figure}[htbp]
\centering\includegraphics[width=10
cm]{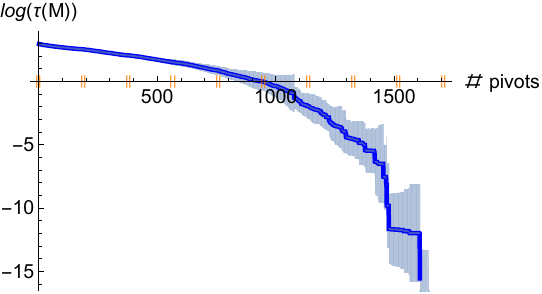}
\caption{Evolution of the average off-diagonal cost function $\log(\tau(M))$ as a function of the pivot count for $d=20$, averaged over fifteen Haar-random unitary matrices. The solid curve represents the ensemble average, while the error bars indicate one standard deviation. The double vertical marks ($||$) denote the completion of successive Jacobi sweeps. The nearly linear decrease during the first sweeps, followed by a progressively faster decay, is characteristic of cyclic Jacobi-type diagonalization algorithms.
 \label{fig2}}
\end{figure}

We first examine the convergence of the JQA under cyclic pivoting. Figure~\ref{fig2} shows the average value of $\log(\tau(M))$ as a function of the pivot count for an ensemble of fifteen Haar-random unitary matrices of dimension $d=20$. Similar behavior was observed for all tested dimensions up to $d=30$.

The off-diagonal cost function decreases monotonically throughout the diagonalization procedure. During the first few Jacobi sweeps, $\log(\tau(M))$ exhibits an approximately linear decrease with the pivot count, consistent with the characteristic convergence of cyclic Jacobi methods. At later stages the convergence accelerates, indicating the onset of the asymptotic quadratic convergence regime observed in the classical algorithm.

\subsection{Scaling of the number of sweeps with the dimension $d$}

An important aspect of JQA is how the number of sweeps required to reach a prescribed accuracy scales with the dimension of the problem. If the observed behavior is to be consistent with an $\mathcal{O}(d^2)$ growth in the number of Jacobi pivot operations, then the number of sweeps should remain approximately independent of the dimension. We recall that a complete sweep consists of $\frac{d(d-1)}{2}$ pivot operations, a quantity fixed by the cyclic pivoting strategy adopted throughout this work.

A second point of interest is to assess the effect of replacing the standard two-parameter Jacobi rotation by the proposed sequence of two single-parameter rotations. Since this decomposition constitutes the main deviation from the classical generic Jacobi algorithm, it is important to verify that it does not adversely affect convergence.

\begin{figure}[htbp]
\centering\includegraphics[width=10
cm]{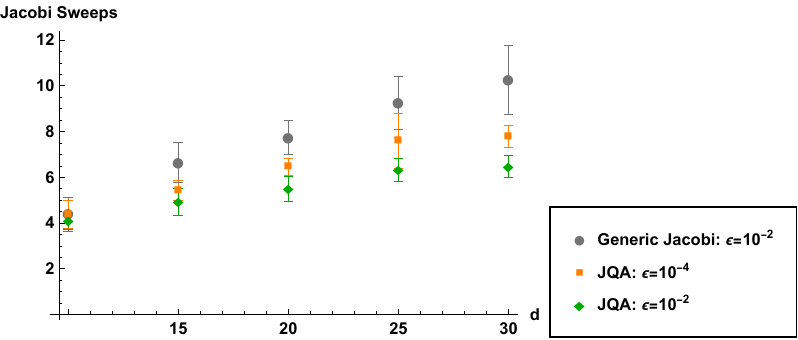}
\caption{Average number of Jacobi sweeps required to reach a target eigenvalue accuracy $\epsilon$ as a function of the dimension $d$. Squares and diamonds correspond to the proposed sequential single-parameter (JQA) implementation for target accuracies $\epsilon=10^{-2}$ and $\epsilon=10^{-4}$, respectively, while circles correspond to the generic Jacobi algorithm for $\epsilon=10^{-2}$. Error bars denote one standard deviation over an ensemble of ten Haar-random unitary matrices. 
 \label{fig3}}
\end{figure}

Both questions are addressed in Fig.~\ref{fig3}. The figure shows the average number of sweeps required to reach the target accuracies $\epsilon=10^{-2}$ and $\epsilon=10^{-4}$, averaged over an ensemble of ten Haar-random unitary matrices. For comparison, results obtained with the generic Jacobi algorithm are also presented for $\epsilon=10^{-2}$.

For both accuracy thresholds, in JQA the average number of sweeps exhibits only a weak dependence on the dimension and stabilizes at approximately six to seven sweeps over the investigated range $10\leq d\leq30$. This observation supports an empirical scaling of
\begin{equation}
N_{\mathrm{pivots}}=
N_{\mathrm{sweeps}}\frac{d(d-1)}{2}
\end{equation}
which is therefore quadratic in the system dimension.

The proposed sequential single-parameter implementation requires a comparable --and in the tested instances slightly smaller--number of sweeps than the standard unitary Jacobi algorithm. Although the improvement is modest, it demonstrates that replacing each two-parameter optimization by two successive one-dimensional searches preserves, and may even slightly improve, the convergence behavior of the original algorithm.

\section{Discussion}\label{discussion}

Having established the convergence and empirical scaling of JQA, we now discuss several aspects of its practical implementation and possible extensions. In particular, we consider restricted connectivity constraints, possible acceleration strategies, extensions to qubit-based architectures, and its potential relevance to early fault-tolerant quantum computing.

\subsection{Restricted set of operations}

The JQA presented in this work, similarly to the classical generic Jacobi algorithm, assumes access to the full set of off-diagonal generators of the Lie algebra $\mathfrak{su}(d)$. In practice, however, experimental platforms often exhibit restricted connectivity, allowing direct implementation of only a subset of the available two-level operations. It is therefore natural to examine the implications of such restrictions for the proposed diagonalization procedure.

Our numerical experiments suggest that, in the absence of additional symmetry assumptions, restricting the available set of generators generally prevents convergence to a fully diagonal matrix. While the off-diagonal norm can still be substantially reduced, a non-vanishing residual typically remains. This behavior is expected, since if a two-dimensional subspace $(p,q)$ cannot be directly addressed, the corresponding off-diagonal matrix elements cannot be independently eliminated within the standard  Jacobi iteration.

From a control-theoretic perspective, this observation reflects the fact that full diagonalization requires access to operations generating the full Lie algebra $\mathfrak{su}(d)$. When direct access to some generators is unavailable, one may in principle attempt to synthesize the missing operations through Lie-algebraic constructions, see for instance \cite{Harel1999}. Such approaches are well known in quantum control theory and can restore controllability of the system. However, the resulting control sequences may become substantially more complex than the elementary two-level operations assumed throughout the present work.

An alternative is to abandon the strict Jacobi pivoting strategy and formulate diagonalization as a global variational optimization over the experimentally available operations. Similar approaches have been proposed for matrix decomposition and eigensystem learning on near-term quantum devices \cite{Wang2021VQSVD}. Whether such a formulation can compensate for restricted connectivity while preserving favorable scaling properties remains an interesting direction for future work.

\subsection{Acceleration strategies and implementation considerations}

Several strategies have been proposed to accelerate the classical Jacobi algorithm. Here we briefly discuss those most relevant to the present quantum implementation.

A widely used acceleration technique is maximal-element pivoting \cite{GolubVanLoan2013}. Instead of following a predetermined cyclic order, one identifies before each pivot the two-dimensional subspace $(p,q)$ satisfying
\begin{equation}  
(p,q)=\arg\max_{i\neq j}\{|M_{ij}|,|M_{ji}|\},
\end{equation}
and applies the Jacobi rotation in that subspace. By always targeting the largest off-diagonal contribution, this strategy typically reduces the number of required sweeps. Within our framework, however, its implementation would require substantial additional measurements, since identifying the maximal off-diagonal element necessitates repeated estimation of many matrix elements, partially defeating the objective of avoiding full matrix reconstruction.

Threshold pivoting \cite{BrentLuk1985} appears more naturally adapted to the present approach. In this technique, Jacobi rotations are applied only to subspaces satisfying
\begin{equation}
|M_{ij}|>\eta
\qquad \text{or} \qquad
|M_{ji}|>\eta,
\end{equation}
where $\eta$ is a prescribed threshold. In this way, pivots that contribute negligibly to the reduction of the off-diagonal norm are skipped. Rather than estimating additional matrix elements, one may instead monitor the optimal angles $\phi_{1,2}$ obtained during the local optimization. If both angles associated with a given pivot remain sufficiently close to zero for several consecutive sweeps, the corresponding operation may be omitted from subsequent sweeps with little impact on convergence.

Another important development is parallel Jacobi diagonalization \cite{BrentLuk1985,DrmacVeselic2008}. The key observation is that generators acting on disjoint two-dimensional subspaces commute. More precisely,
\begin{equation}
[\hat g_{(p,q)},\hat g_{(r,s)}]=0,
\end{equation}
whenever the pairs $(p,q)$ and $(r,s)$ do not share a common index. Consequently, a sweep can be partitioned into subsets of mutually commuting operations that may be executed simultaneously on independent processors.  The same idea could be explored on quantum hardware by executing commuting rotations on independent QPUs or qudit processors, thereby potentially reducing the effective circuit depth.

Finally, Block Jacobi methods \cite{Hari1995} replace the elementary $2\times2$ pivots  by larger $k\times k$ blocks, while the associated $SU(2)$ rotations are replaced by  $SU(k)$ transformations. Although involving more variables, the local optimization becomes more demanding, several pivots are treated simultaneously, often reducing the number of sweeps required for convergence. From the perspective of the present work, Block Jacobi naturally interpolates between the highly structured Jacobi procedure and fully variational matrix decomposition methods. In the limiting case $k=d$, the distinction between local pivot optimization and global variational optimization effectively disappears.

For clarity, we deliberately adopted cyclic pivoting throughout this work. Although not the fastest classical variant, cyclic pivoting requires no additional matrix information, is deterministic, possesses well-understood convergence properties, and maps naturally onto the proposed experimental implementation.

\subsection{From qudit to qubits}

A natural question  is whether JQA can be implemented on a register of $n$ qubits when $d=2^n$. Since the elementary building blocks of the algorithm are $SU(2)$ rotations acting on two-dimensional subspaces of a $d$-dimensional Hilbert space, to this end we need to examine how such operations translate into a qubit representation.

To illustrate the complexity of such a mapping, we just consider the simplest non-trivial case of a three-qubit register corresponding to $d=8$. The first three generators $\hat{g}^{(R)}_{(p,q)}$ can be expressed,  as linear combinations of Pauli strings acting on the qubits:
\begin{eqnarray}
   \hat{g}^{(R)}_{(2,1)} &=&\frac{1}{4}(\hat{Z}_1\hat{Z}_2\hat{X}_3+\hat{Z}_2\hat{X}_3+\hat{Z}_1\hat{X}_3+\hat{X}_3) \\
   \hat{g}^{(R)}_{(3,1)} &=&\frac{1}{4}( \hat{Z}_1\hat{X}_2\hat{Z}_3+\hat{X}_2\hat{Z}_3+\hat{Z}_1\hat{X}_2+\hat{X}_2)\\
    \hat{g}^{(R)}_{(4,1)} &=&\frac{1}{4}(\hat{Z}_1\hat{X}_2\hat{X}_3+\hat{X}_2\hat{X}_3-\hat{Z}_1\hat{Y}_2\hat{Y}_3-\hat{Y}_2\hat{Y}_3).\\
\end{eqnarray}
where $\{\hat{X}_i,\hat{Y}_i,\hat{Z}_i\}$ denote the Pauli operators acting on the  $i$-th qubit.

This example clearly exhibits   that the qudit generators appearing naturally in the proposed construction generally correspond to non-trivial combinations of multi-qubit operators when represented in a qubit basis. Consequently, while the algorithm can in principle be implemented on qubit architectures, the elementary qudit rotations must first be compiled into sequences of native qubit gates. The resulting circuit therefore inevitably inherits the overhead associated with decomposing higher-dimensional $SU(d)$ operations into elementary one- and two-qubit gates, as also illustrated by previous qubit implementations of Jacobi-type algorithms \cite{Parrish2019}.

\subsection{Relevance for fault tolerance}

The proposed Jacobi qudit algorithm possesses  several features that may prove attractive for early fault-tolerant architectures. The method avoids controlled-unitary constructions and ancilla-assisted phase estimation, relying instead on elementary two-level rotations and measurements of a single observable. Furthermore, its highly structured form, inherited from the classical Jacobi procedure, makes the required operations transparent. Since independent Jacobi pivots acting on disjoint subspaces commute, the algorithm also appears naturally suited to parallel and distributed quantum implementations, where commuting rotations could be executed simultaneously on independent QPUs or qudit processors.

The inherently qudit-native formulation provides an additional motivation. Rather than encoding the problem into qubits from the outset, the algorithm exploits the structure of the $\mathfrak{su}(d)$ algebra directly through elementary two-level transformations acting on a single qudit. Together with recent progress in qudit-based quantum error correction \cite{Brock2025}, these features suggest that structured Jacobi-type diagonalization may provide an attractive alternative to more circuit-intensive eigensystem extraction methods in early fault-tolerant quantum computing.

\section{Conclusions}\label{conclusions}

In this work, we have proposed a structured quantum implementation of the Jacobi diagonalization algorithm for unknown unitary operators acting on a single qudit. JQA follows the philosophy of the classical generic Jacobi algorithm while replacing the identification of each elementary two-parameter Givens rotation by two sequential single-parameter variational searches. Numerical experiments on ensembles of Haar-random unitary matrices indicate that this modification preserves the characteristic convergence properties of the classical method while exhibiting comparable, and in the tested instances slightly improved, convergence behavior.

The objective of the present work is not to solve a classical diagonalization problem by means of a quantum computer. Such an approach would inevitably suffer from the overhead of encoding a classical matrix into a quantum circuit, while the algorithmic complexity remains comparable to that of the classical Jacobi algorithm. Instead, the proposed method is intended for the direct diagonalization of unknown quantum operations or Hamiltonians that can be accessed experimentally but are not known classically. In this setting, the algorithm avoids reconstrucing the operator and simultaneously provides both the eigenvalues and the associated eigenvectors of the operator.

Another attractive feature of JQA is that it naturally inherits several characteristics of the classical Jacobi algorithm. In particular, the possibility of parallel execution, which has recently renewed interest in Jacobi methods within classical high-performance computing, can in principle also be exploited in quantum architectures by executing commuting rotations on independent qudit processors. Although the practical realization of such distributed implementations remains an open question, the structured nature of the algorithm appears particularly well suited to future parallel quantum hardware.

The discussion presented in this work also identifies several directions deserving further investigation. These include the incorporation of classical acceleration strategies, such as threshold or block Jacobi methods, a more rigorous convergence analysis of the sequential optimization procedure, and the treatment of restricted-connectivity architectures. Another interesting theoretical direction is the study of qubit realizations of the JQA. However, as discussed in Sec.~\ref{discussion}, the elementary qudit operations generally require non-trivial compilation into multi-qubit circuits, introducing a substantial overhead that largely removes the natural advantages of the qudit formulation. Consequently, such an extension should primarily be viewed as a means of understanding the relationship between qudit and qubit implementations, rather than as a practical route for realizing the proposed algorithm.

More broadly, we believe that the interaction between classical numerical linear algebra and quantum algorithm design offers a fertile research direction. Classical algorithms developed over many decades can inspire new structured quantum procedures, while quantum implementations may, in turn, motivate new variants of their classical counterparts. The slight improvement observed here by decomposing each two-parameter optimization into two sequential one-dimensional searches provides one example of this mutually beneficial exchange. We hope that the present work contributes to this growing dialogue and stimulates further investigation of structured quantum numerical algorithms.



\bibliographystyle{unsrturl}
\bibliography{references}

\end{document}